\documentclass[referee]{mn2e}
\usepackage{graphics}
\title[Polycrystalline graphite model for the 2175 \AA{\ } band]{A polycrystalline graphite model for the 2175 \AA{\ } interstellar extinction band}
\author[R. J. Papoular and R. Papoular]{Robert J. Papoular$^{1}$\thanks{E-mail:
Robert.Papoular@cea.fr} and 
Renaud Papoular$^{2}$\thanks{E-mail: papoular@wanadoo.fr}\\
$^{1}$IRAMIS, Laboratoire Leon Brillouin, CEA Saclay, 91191 Gif-s-Yvette, France\\
$^{2}$Service d'Astrophysique and Service de Chimie Moleculaire,
 CEA Saclay, 91191 Gif-s-Yvette, France}
\begin{document}

\date{Accepted . Received ; in original form }

\pagerange{\pageref{firstpage}--\pageref{lastpage}} \pubyear{2002}

   \maketitle
\label{firstpage}
\begin{abstract}
A random, hydrogen-free, assembly of microscopic sp$^{2}$ carbon chips, forming a macroscopically homogeneous and isotropic solid, is proposed as a model carrier for the UV interstellar extinction band . The validity of this model is based on the calculation of the Bruggeman average dielectric function of a mixture of the known parallel and perpendicular dielectric functions of graphite. The $\pi$ absorption feature of Rayleigh-sized spheres of this mixture falls near 4.6 $\mu$m$^{-1}$ 
(2175 \AA{\ }), but its width is 1.5 $\mu$m$^{-1}$, somewhat larger than the astronomically observed average, 1 $\mu$m$^{-1}$. This is confirmed by measurements of the reflectance of an industrial material, polycrystalline graphite. A better fit to the IS feature position and width is obtained with a hypothetical material, having the same dielectric functions as natural graphite, except for less extended wings of the $\pi$ resonance. Physically, this could result from changes in the electronic band structure due to previous thermal histories. On this model, the Fr\"olich feature central wavelength depends only on the $\pi$ resonance frequency, while its width depends only on the damping constant of the same resonance. This explains the range of observed feature widths at constant feature wavelength.
  \end{abstract}
\begin{keywords}
Stars:Astrochemistry; interstellar matter; ISM: dust; molecular processes.
\end{keywords}

\section{Introduction}
The numerous carbon models that have been proposed to mimic the 2175 \AA{\ } (or 4.6 $\mu$m$^{-1}$) interstellar (IS) extinction band can roughly be divided into two generic groups respectively characterized by their ordered or disordered solid structure. Typical of the first group is the distribution of randomly oriented pure graphite spheres (see Draine and Lee \cite{dl84}). Fullerenes and bucky onions (BO) were later proposed (e.g. De Heer and Ugarte \cite{deh}, Wada and Tokunaga \cite{wad}). These recently returned to the forefront, with the laboratory isolation of single BOs by Chhowalla et al. \cite{chh}; their spectral extinction was measured and shown to exhibit a band at 4.55$\pm0.1 \mu$m$^{-1}$ whose profile, when adequately smoothed, fits the ``Drude-like" profile proposed by Fitzpatrick and Massa \cite{fm}. This experimental result is backed by theoretical analyses, one of the latter being that of Ruiz et al. \cite{rui}, who demonstrated a very good agreement of their result with the measurement.

The main difficulties faced by such ``ordered" models are (see Draine and Lee \cite{dl84}, Draine and Malhotra \cite{dm93}): 
 
a) the absence of a known process by which the FWHM of the band could vary by a factor of $\sim$2, as observed, even as its peak and shape remain unaltered;

b) in particular, the measured range of FWHMs of BOs is 1.2-1.6 $\mu$m$^{-1}$, outside the observed range (0.7 to 1.3 $\mu$m$^{-1}$); it is not clear how BOs could be altered to yield FWHMs in the required range;

c) the appearence, in the spectrum of the model, of several features that are not observed in the sky, e.g. the four IR features of the C$_{60}$ molecule (see Huffman \cite{huf});

d) the absence of efficient IS processes capable of producing enough fullerenes or BOs, which are known to be hard to obtain, even in specialized laboratories (see Chhowalla et al. \cite{chh}). The same applies to pure graphite particles.

These difficulties prompted research in the opposite direction, namely amorphous carbon. This ill-defined term covers a large family of materials, hypothetized or indeed known in the laboratory (see Robertson \cite{rob}), and characterized by an array of $\pi$ and $\sigma$ bonds randomly distributed in space, on the \emph{microscopic} scale. This is not to be confused with random \emph{macroscopic} clusters of graphitic particles (e.g. Draine \cite{dra88}), or composite grains (e.g. Mathis and Whiffen \cite{mat}). One of the first disordered carbon model for the 2175 \AA{\ } band was proposed by Sakata et al. \cite{sak}) who synthesized a quenched carbonaceous composite material (QCC), displaying a UV absorption near 2000 \AA{\ }. A number of workers followed suit, (for references, see Bussoletti et al. \cite{bus}. Common to these workers is, apparently, the idea explicited by Hecht \cite{hec}, based on the work of Donn \cite{don} and Nuth \cite{nut} on kerogen-like models of IS grains, and the evolutionary trend of such earthly materials upon heating, as observed, in particular, by Fink et al. \cite{fin}: it is suggested that, in this graphitization process, the H atoms they initially contain are progressively expelled while the essentially C residual atoms are spatially redistributed, allowing somehow the 2175 \AA{\ } feature to emerge.

Pursuing this effort, we return here to our polycystalline graphite (PG) model (Papoular et al. \cite{pap93}, using a new approach and bringing new evidence in favour of such a model. In Sec. 2, we first show on general grounds, how the position and width of the band can be disconnected for an amorphous material, and what their governing factors are.

 As an embodiment of this notion, we envision in Sec. 3 an isotropic homogeneous material made up of randomly oriented, atomic-scale, graphitic lumps; its dielectric function is computed, according to the Bruggeman rule, by mixing the lumps in adequate concentrations. This is shown, in the Rayleigh approximation of small nearly spherical grains, to yield a strong band at 20 \AA{\ } away from the observed IS wavelength, with a FWHM of 1.6 $\mu$m$^{-1}$.

 We then tackle the problem of reducing the band width to the IS standards without letting the central wavelength drift excessively: this is done by varying the dielectric properties of the constituent ``graphite" only to a very limited extent. In this process, the band retains a shape that reasonably fits the ``Drude-like" profile and its intensity satisfies the IS C-abundance constraint. 

 Section 4 compares these results with our previous measurements of man-made PG (Papoular et al. \cite{pap93}): it is shown that the latter are adequately mimicked by our model associated with common graphite, lending more credibility to the ``amorphous" model. This provides a link to models using less graphitized constituent materials, which may help explaining the many observed ``red" outlier bands. Finally, Section 5 discusses in more detail the compatibility of the model with observational constraints.

\section{A property of the Fr\"ohlich resonance}

The weak scattering apparently included in the UV band extinction, elicits the assumption of small particles for the band carrier, and, hence, the application of the Rayleigh approximation. As a corollary, surface effects cannot be neglected, which entails the use of the Fr\"{o}hlich expression for the extinction efficiency of a small sphere of radius $a$:

\begin{equation}
Q/a=\frac{24 \pi}{\lambda}\frac{\epsilon_{2}}{(\epsilon_{1}+L^{-1}-1)^{2}+(\epsilon_{2})^{2}}
\end{equation}

where $\epsilon_{1,2}$ are, respectively, the real and imaginary parts of the dielectric function of the assumed homogeneous material; $L$ is the shape parameter of the supposedly ellipsoidal grain, 1/3 in the case of a sphere. We assume throughout that scattering is negligible so that there is no need to distinguish between absorption and extinction efficiencies, both being designated by $Q$.

In order to quantitatively make our point, we focus on a spectral region where the dielectric function of the material can be represented by a Lorentzian oscillator (Bohren and Huffman \cite{boh}) with resonant frequency $\omega_{0}$, plasma frequency $\omega_{p}$ and damping constant $\gamma$ (assumed small):

\begin{equation}
\epsilon_{1}(\omega)=1+\frac{\omega_{p}^{2}(\omega_{0}-\omega)/2\omega_{0}}{(\omega_{0}-\omega)^{2}+(\gamma /2)^{2}}; \,\,
\epsilon_{2}(\omega)=\frac{\omega_{p}^{2}\gamma/4\omega_{0}}{(\omega_{0}-\omega)^{2}+
(\gamma /2)^{2}}.
\end{equation}

Plots of $\epsilon_{1,2}$ and $Q/a$ for different values of the Lorentzian parameters show that, there being spectral dispersion, the maximum of $Q$ does not occur when
$\epsilon_{1}=1-L^{-1}$, which is true only if $\gamma/\omega_{0}$ is vanishingly small; $\epsilon_{1}$ need not even fall below  $1-L^{-1}$ for a Fr\"{o}hlich resonance to occur. In the case of graphite, contrary to crystals like SiC, for instance, the excursion from this equality is considerable and depends on the slope of $\epsilon$ vs frequency. Inspection of the plots suggests instead that

\begin{equation}
\epsilon_{1}+2=\epsilon_{2},
\end{equation}

 The \emph{numerical} fact that this relation is approximately verified near the maximum of $Q/a$ can be understood by noting that, in the vicinity of our resonance, $\epsilon_{1,2}$ both vary, but in opposite senses. Since they appear in a sum of two terms in the denominator of $Q/a$ (eq. 1) it seems reasonable to assume, as an approximation, that the maximum (at $\omega_{m}$) occurs when the two terms are equal.

Relation 3 is the more closely verified the smaller $\gamma$ is with respect to $\omega_{0}$ and $\omega_{p}$. If it is not too large, one readily concludes that

\begin{equation}
\omega_{m}\sim\omega_{0}+\frac{\omega_{p}^{2}}{6\omega_{0}}.
\end{equation}
For an ellipsoid in general, the factor 6 in the denominator must be replaced with $2/L$, meaning that the small variations of the Fr\"olich resonance frequency are nearly linear in $L$. Note the absence of $\gamma$ in this expression. Now to evaluate the FWHM of the resonance. By definition, when $\omega_{1/2}=\omega_{m}\pm$FWHM/2, 

\begin{equation}
Q/a=Q_{max}/2a=\frac{24 \pi}{\lambda}\frac{1}{2\epsilon_{2}(\omega_{m})}.
\end{equation}

Rewriting this expression in terms of the 3 Lorentzian parameters, and assuming again that $\gamma<<\omega_{0,p}$, one obtains

\begin{equation}
\omega_{1/2}\sim\omega_{m}\pm\frac{\sqrt 3\gamma}{2}\,;\,\, FWHM=\sqrt 3\gamma .
\end{equation}

Thus, the sharpness of the Fr\"ohlich resonance increases linearly with the steepness of the Lorentzian (blue) wing. Here, only $\gamma$ appears. That the dependencies of FWHM and peak frequency are distinct is not specific of the Lorentzian but results from the Kramers-Kronig link between the real and imaginary parts of the dielectric function. This result will help, below, tailoring the width of the model UV extinction band.

\section{The model carrier} 
It is now widely recognized that IS carbon grains cannot be pure graphene chunks, but also include ``impurities" (mainly hydrogen, oxygen and nitrogen) which destroy the beautiful symmetry of the structure, thus determining the direction of the subsequent evolutionary trend. The energy absorbed by the grain in the harsh IS medium progressively removes the volatile impurities, which are less tightly bound than carbon. Starting from a disordered state, this process can only add disorder to the structure. Our model carrier is in the ultimate stage of this evolution, where, in order to obtain a $\pi$-type resonance, we have to assume that all  adatoms have been expelled, and the carbon bonds are mainly of the sp$^{2}$ type, like in polycrystalline graphite or glassy carbon (Taft and Philipp \cite{taf}, Robertson \cite{rob}). More precisely, this can be construed as an aggregate of small stacks of graphene planes which, instead of piling up one upon the other, normally to their common direction, would lie at a random angle to one another. This arrangement is inspired by the structure first described by Franklin \cite{fra} for ``non-graphitizing", later called glassy carbon. 

 The size of each stack is too small for the stacks to be considered as ellipsoids to which Fr\"ohlich's theory may be applied. Instead, we consider that the constituent material of the grain is microscopically homogeneous, with electromagnetic properties resulting from an adequate mix of two different materials. Indeed, along any given direction, each stack is viewed either approximately parallel 
or approximately perpendicular to the graphene planes. It is assumed here that the situation is akin to that of a mixture of two homogeneous and \emph{isotropic} materials, whose dielectric functions are, respectively, $\epsilon\parallel$ and $\epsilon\bot$ of graphite, with corresponding concentrations of 1/3 and 2/3. It is usually considered that, for such mixtures, the Bruggeman mixing formula  is a good choice among the host of proposed formulae (Bohren and Huffman \cite{boh}). In the present instance, we write this formula as

\begin{equation}
f_{\parallel}\frac{\epsilon\parallel-\epsilon_{eff}}{\epsilon\parallel+2\epsilon_{eff}}
+(1-f_{\parallel})\frac{\epsilon\bot-\epsilon_{eff}}{\epsilon\bot+2\epsilon_{eff}}=0 \,\,,
\end{equation}

where $f_{\parallel}=1/3$ and $\epsilon_{eff}$ is the required isotropic, spatially uniform dielectric function. The factor 2 in the denominators is the particular value of a function which depends on the geometrical factor, $L$, defining an ellipsoid, and is valid for spheres ($L=1/3$). This procedure has recently been rediscussed (Lucarini et al. \cite{luc}), and  shown to satisfy the Kramers-Kronig causality relationship.

It must be stressed that, for complex dielectric functions, the Bruggeman equation admits of several solutions, among which a physically sensible choice must be made. A copy of the code we have written can be obtained upon request. For $\epsilon\parallel$, we used the tabulated values in Draine \cite{dra85}; for $\epsilon\bot$, we used both the original values of Taft and Philipp \cite{taf} and those adjusted by Draine and Lee \cite{dl84} to deliver the right UV band according to their model of randomly oriented pure graphite spheres. Figure 1 shows that the Draine and Lee values lead to a somewhat better overall band shape, when compared with the fittest ``Drude" profile of Fitzpatrick and Massa \cite{fm}.

\begin{figure}
\resizebox{\hsize}{!}{\includegraphics{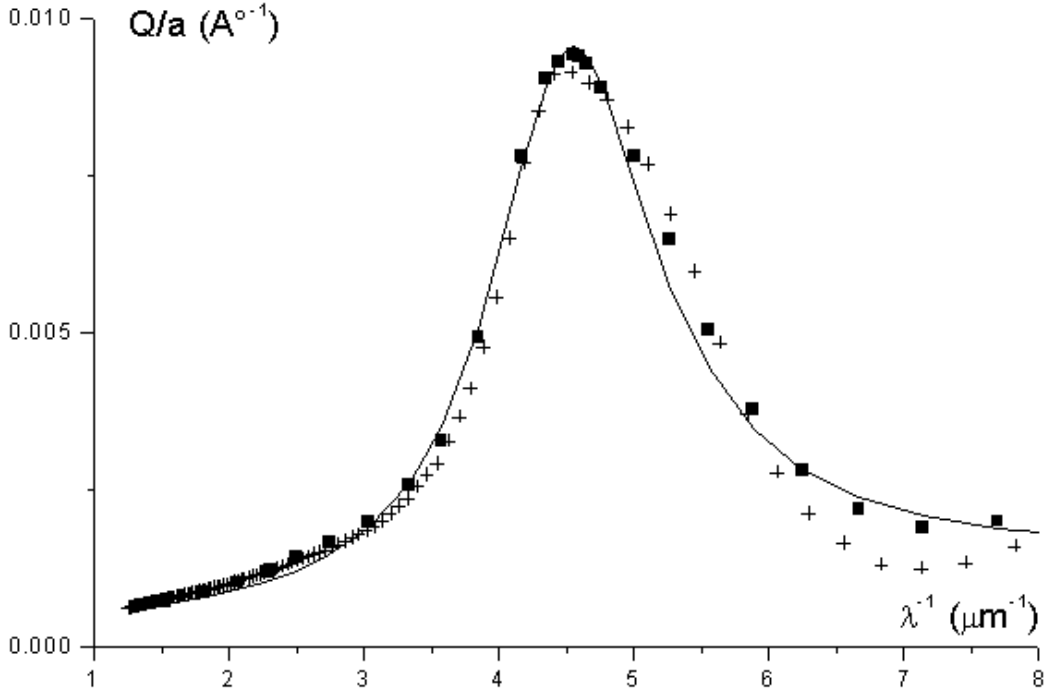}}
\caption[]{The UV extinction feature deduced from the Bruggeman mixing formula applied to $\epsilon\parallel$ and $\epsilon\bot$ (respectively 1/3, 2/3). Squares: $\epsilon\parallel$ and $\epsilon\bot$ from Draine \cite{dra85}; pluses: $\epsilon\bot$ from Taft and Philipp \cite{taf}; line: ``Drude" profile with $\lambda_{0}$=2200 \AA{\ }, $\gamma=1.5\, \mu$m$^{-1}$.}
\end{figure}

Both features peak at 2200 \AA{\ }, near the observed wavelength; however they are both wider (1.5  $\mu$m$^{-1}$) than the astronomically observed average, 1 $\mu$m$^{-1}$. Some tailoring of the dielectric properties is therefore in order, and Sec. 2 may help in doing this.
 
According to eq. 4, a reduction of only $\sim3\%$ in the $\pi$ plasma frequency is enough to bring the extinction peak frequency to the desired value. Such a change is conceivable in space, as  $\omega_{p}$ is intimately linked to the electrical conductivity, which is very sensitive to the heat treatment previously suffered by the material (Robertson \cite{rob}). Alternatively,
the feature location may be controlled through $\epsilon\parallel$, which is notoriously subject to experimental errors of measurement (see Jellison et al. \cite{jel}) and was recently shown to strongly depend on the distance between graphene planes (Fei et al. \cite{fei}, Marinopoulos et al. \cite{mar}). Thus, calculations show that changing its value by a factor of 1/2 or 2 shifts the feature to 2070 and 2210 \AA{\ }, respectively; the required adjustment may thus fall within the limits of measurement errors and structural variations. 

As for the band width, eq. 6 shows that it can be reduced by steepening the blue wing of the $\pi$ resonance. Now, solid state physicists have shown that the Lorentzian representation of the dielectric function is suited to cases where a single decay channel for excitation exists or is dominant. 

On the other hand, the Gaussian representation, which results in a quicker evanescence of the wings, is known to be preferable when the decay is primarily associated with a large number of weak scattering centers (Garland et al. \cite{gar} and references therein; Franke et al. \cite{franke}), which is likely the case for our disordered model. It was used, for instance, by Brendel and Bormann \cite{bre} for amorphous solids.

 One combination of Lorentzian and  Gaussian representations is the so-called pseudo-Gaussian (Kim \cite{kim}), obtained by replacing the constant $\gamma$ in the Lorentzian by a function of the form

\begin{equation}
\gamma(\omega)=\frac{\gamma_{0}}{exp[\alpha((\omega-\omega_{0})/\gamma_{0})^2]}.
\end{equation}

This was found to be helpful, in particular, for terrestrial graphite (Djurisic and Li \cite{dju}). Here, we used the latter device to represent the $\pi\bot$ resonance, and Lorentzians for the other 3 resonances. $\epsilon$// and $\epsilon\bot$ are each the sum of the corresponding $\sigma$ and $\pi$ functions. Table 1 lists all the constants used to compute 3 band spectra, covering a representative range of FWHM widths; the last entry, $\gamma_{\pi\bot}$, is given by eq. 8, where $\gamma_{0}=2\,10^{15}$ rad.s$^{-1}$ and $\alpha$ was given 3 values, successively: 1/4, 1, 4; only $\alpha$ was changed in the process, not $\omega_{0}$, nor $\omega_{p}$, so as to limit the excursion from natural graphite properties to a minimum. Figure 2 illustrates the results.

\begin{table*}
\caption[]{Oscillator parameters}
\begin{flushleft}
\begin{tabular}{lllllll}
\hline
Polarization & & $\epsilon\parallel$ & & & $\epsilon\bot$ &  \\
\hline
Resonance & $\omega_{0}$ & $\omega_{p}$ & $\gamma$ & $\omega_{0}$ & $\omega_{p}$ & $\gamma$ \\
\hline
$\sigma$ & $1.7\,10^{16}$ & $1.8\,10^{16}$ & $3\,10^{15}$ & $2.17\,10^{16}$ & $2.8\,10^{16}$ & $4\,10^{15}$ \\
\hline
$\pi$ & $5.94\,10^{15}$ & $8.5\,10^{15}$ & $4\,10^{15}$ & $6.73\,10^{15}$ & $1.12\,10^{16}$ & $\gamma_{\pi\bot}$ \\
\hline
All values in rad.s$^{-1}$; $\gamma_{\pi\bot}$ from eq. 8.
\end{tabular}
\end{flushleft}
\end{table*}

\begin{figure}
\resizebox{18cm}{21cm}{\includegraphics{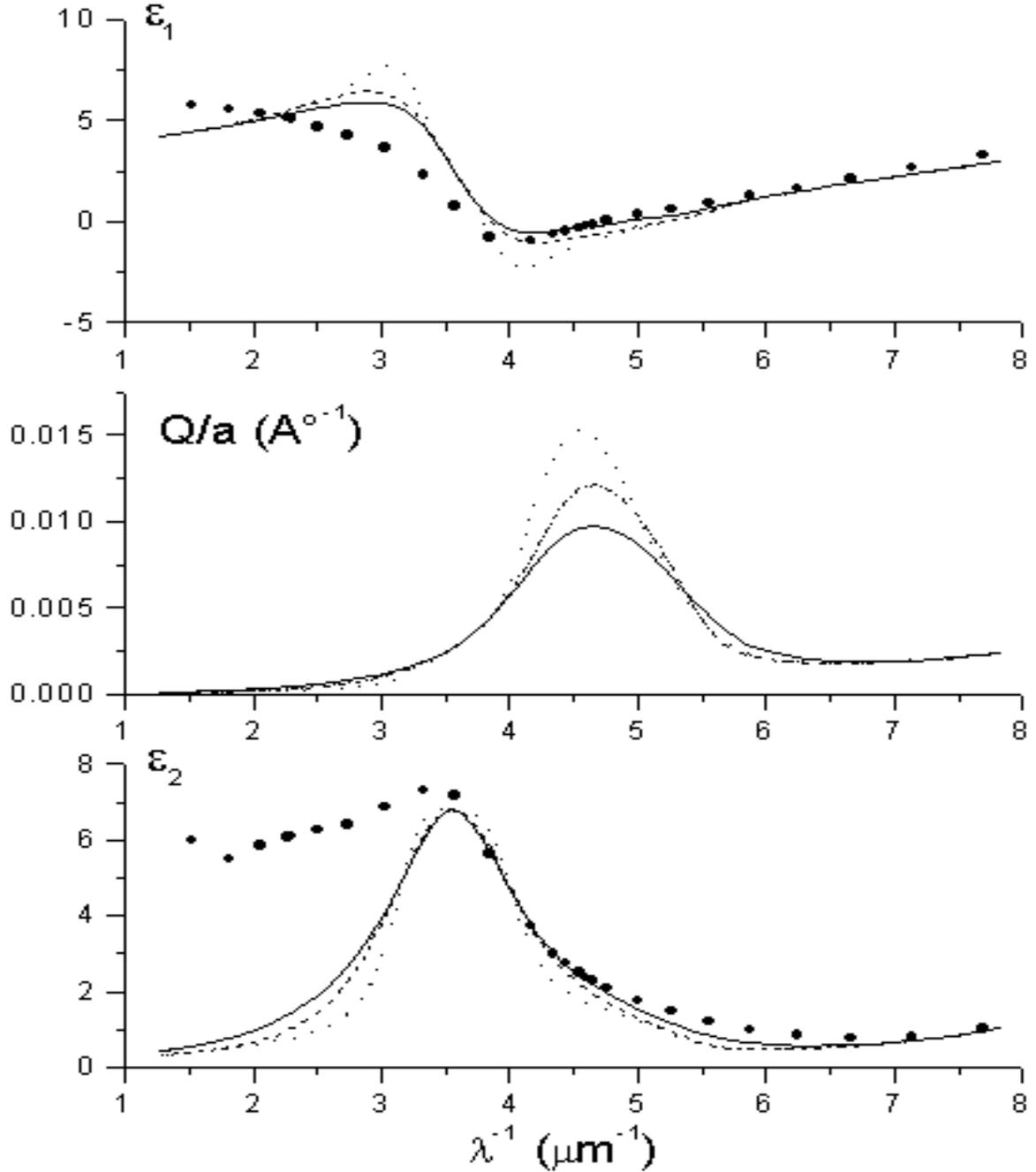}}
\caption[]{Middle graph: $Q/a$ (\AA{\ }$^{-1}$) of a sphere of radius $a$, made of the model carrier (Sec. 3), in the spectral window of interest. $\epsilon_{1}$ and $\epsilon_{2}$ in eq. 1  were obtained from the Bruggeman averaging formula (7), successively applied to 3 pairs of Lorentzian-like $\epsilon\parallel,\bot$ (see text and Table 1). Only one parameter, $\alpha$, characterizing the width of the $\pi\bot$ resonance varies from case to case; $\alpha$=0.25 (line), 1 (dashes), 4 (dotted), respectively. The extinction feature grows and narrows as $\alpha$ increases.
  Upper and lower graphs: the 3 corresponding $\epsilon_{1}$ and $\epsilon_{2}$. For comparison, the heavy dots represent the average dielectric functions obtained from the Bruggeman formula by inserting the dielectric functions tabulated by Draine \cite{dra85} for $\epsilon\parallel$ and $\epsilon\bot$ of ``astronomical" graphite.}
\end{figure}

In Fig. 2, the upper and lower graphs, respectively, plot $\epsilon_{1}$ and $\epsilon_{2}$ for the 3 values of $\alpha$: 1/4, 1, 4, as well as the corresponding curves for a Bruggeman mixture $\epsilon_{\parallel}/3+2\epsilon_{\perp}/3$, taken from Draine \cite{dra85}. Note the limited extent of differences between the synthesized dielectric functions of the 3 cases, in the UV band window, 3-6 $\mu$m$^{-1}$. Below 4 $\mu$m$^{-1}$, most of this difference is due to the simplistic representation of the material by only 2 pairs of Lorentzian oscillators. The middle graph represents the corresponding extinction efficiencies in the Rayleigh limit.
Fitting a ``Drude" function to each delivers the width and corresponding central wavelength

$\gamma$=1.6, 1.35, 1 $\mu$m$^{-1}$

$\lambda_{0}$=2150, 2150, 2190 \AA{\ }.

Figure 3 illustrates the trends of $\gamma$ and $\lambda_{max}$ as $\alpha$ varies. In agreement with Fig. 2 of Fitzpatrick and Massa (1986), $\lambda_{max}$ remains within 1.1 $\%$ of 2175 \AA{\ }.

\begin{figure}
\resizebox{\hsize}{!}{\includegraphics{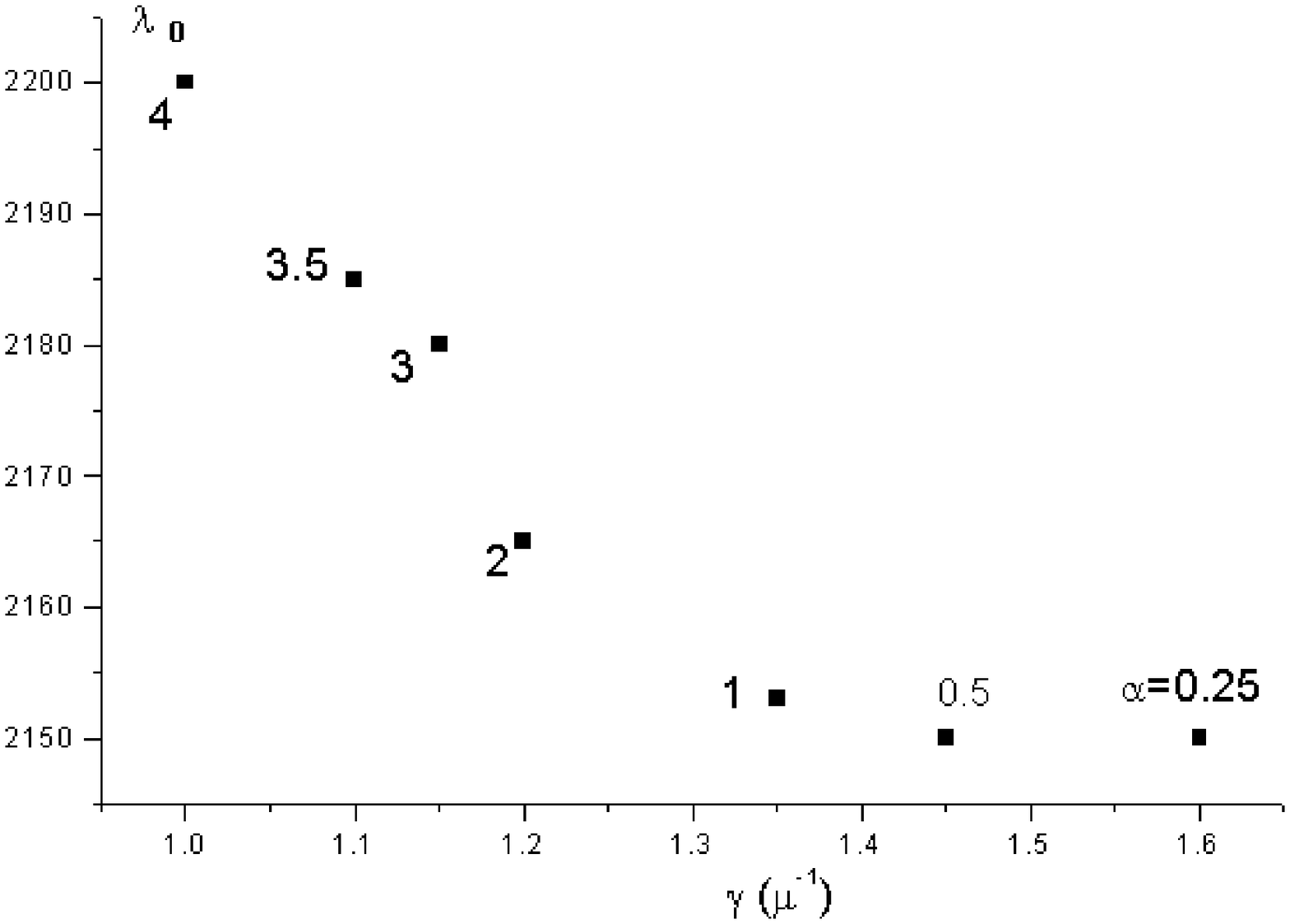}}
\caption[]{The trends of $\gamma$ and $\lambda_{0}$ as $\alpha$ varies (eq. 4 and 6). In agreement with Fig. 2 of Fitzpatrick and Massa (1986), $\lambda_{max}$ remains within 1.1 $\%$ of 2175 \AA{\ }.}
\end{figure}

Figure 4 shows the Drude fit for the narrowest feature.

\begin{figure}
\resizebox{\hsize}{!}{\includegraphics{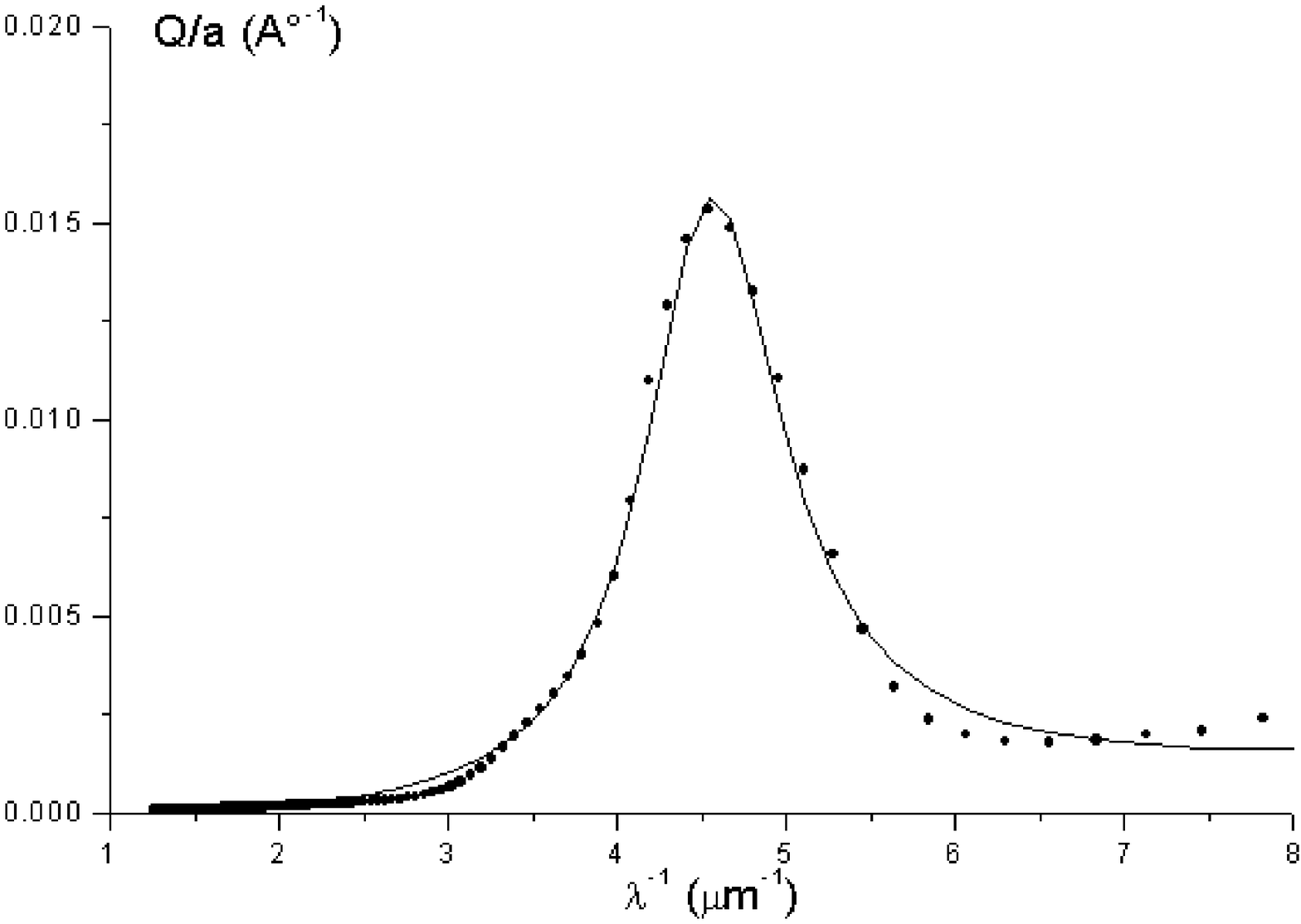}}
\caption[]{``Drude" fit (line) for the narrowest feature of Fig.2 (dots): $\lambda_{0}=2190$ \AA{\ }, $\lambda^{-1}=4.566\,\mu$m$^{-1},\,\,\gamma=1\,\mu$m$^{-1}$}
\end{figure}

Even though this fit may be considered as acceptable, it is limited to a small wavenumber interval. Based on the astronomical observations in this interval alone, one cannot infer the overall properties of the carrier. However, the variations of the dielectric properties, required to cover the diversity of the observed features, are  limited, and the difference is small between these properties and those of ``astronomical" graphite. More quantitatively, the effective electron number as a function of frequency, as deduced from the Sum Rule (see Taft and Philipp \cite{taf}) for the cases considered in Fig. 2 is hardly distinguishable from that of astronomical graphite. A reasonable surmise is, therefore, that the latter may be taken as a first approximation to the functions of our model carrier.

\section {A laboratory material embodiment: Polycrystalline graphite (PG)}

High purity PG is produced industrially for various purposes. It is not to be confused with HOPG, or Highly Oriented Pyrolytic Graphite, which is also produced under high temperatures and high pressures. HOPG is very close to the perfect natural graphite, while PG is perhaps the best yet embodiment of the opposite extreme: a random assembly of microscopic sp$^{2}$ carbon chips pressed together, (also termed ``turbostratic crystallites" in the Carbon literature; see Mrozowski \cite{mro}), which we propose for a model band carrier. We have measured the normal-incidence reflectance of the polished surface of a pressed pellet made from high-purity PG powder (Papoular et al. \cite{pap93}).  This provides us now with the opportunity to test the validity of the theoretical mixing procedure presented above. Figure 5 displays the measured reflectance of PG, which we now set out to fit with the model of Sec. 3. For this purpose, we follow the same procedure, and define 4 oscillators as above. Since the material is supposedly pure graphite, we try first the same oscillators as in Table 1, except that, for a start, we adopt a constant $\gamma$ for all 4 oscillators. Trying different values of $\gamma(\pi\perp)$, all other parameters remaining unchanged, we compute the dielectric function of the mixture ``$\epsilon\parallel/3,\, 2\epsilon\bot/3$", using the Bruggeman formula, then derive the corresponding optical indices, $n$ and $k$, using the following formulae:

$n=(\frac{(\epsilon_{1}^{2}+\epsilon_{2}^{2})^{1/2}+\epsilon_{1}}{2})^{1/2}$

$k=(\frac{(\epsilon_{1}^{2}+\epsilon_{2}^{2})^{1/2}-\epsilon_{1}}{2})^{1/2}$.

The normal reflectance is then given by 

$R=\frac{(n-1)^{2}+k^{2}}{(n+1)^{2}+k^{2}}$.

\begin{figure}
\resizebox{\hsize}{!}{\includegraphics{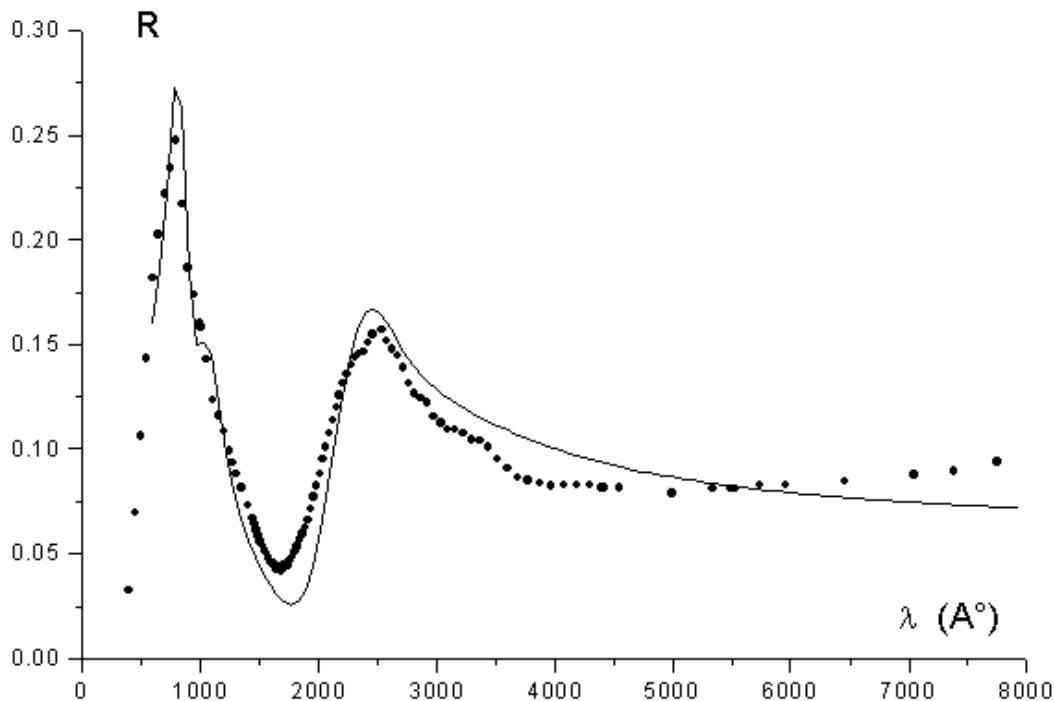}}
\caption[]{Dots: normal incidence reflectance of polished pressed pellet of industrial polycrystalline graphite powder (PG) measured at the LURE synchrotron; line: PG model of Sec. 3, with the same parameters as for the narrow feature of Fig. 4, except $\gamma(\pi\bot)=2\,10^{15}$ rad.s$^{-1}$, constant; $\omega_{p}=7.48\,10^{15}$ rad.s$^{-1}$, giving an approximate fit, which shows the ability of the model to mimic a laboratory material approaching the ideal random sp2 structure.}
\end{figure}

\begin{figure}
\resizebox{\hsize}{!}{\includegraphics{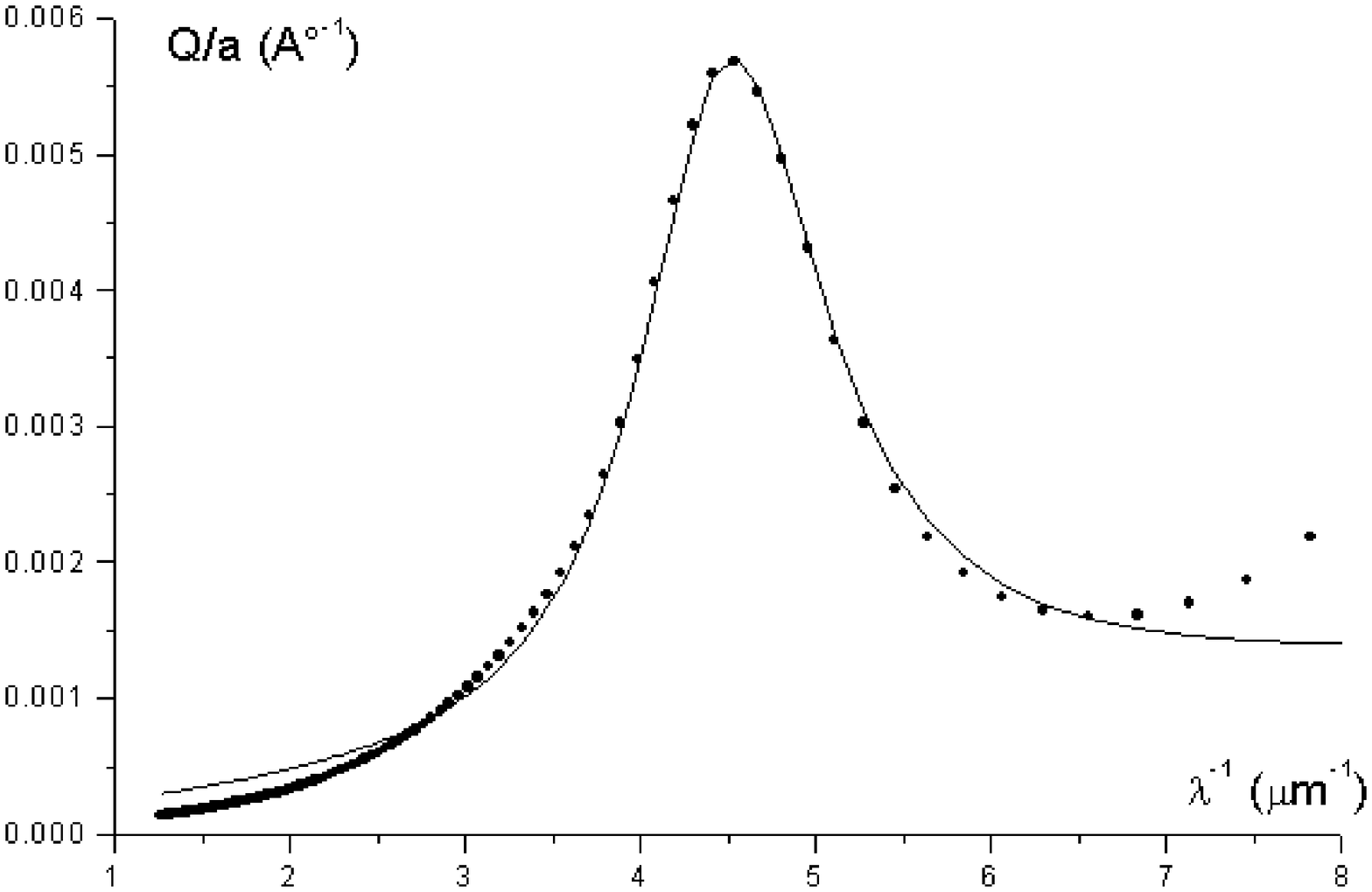}}
\caption[]{The UV extinction feature of the laboratory PG of Fig. 5 (dots), together with a ``Drude" fit, with  $\lambda_{0}=2210$ \AA{\ }, $\gamma=1.3\,\,\mu$m$^{-1}$.}
\end{figure}
It is found that the closest fit is obtained with $\gamma=constant=2\, 10^{15}$ rad.s$^{-1}$; but still a  better fit results from a minor change in Table 1, namely $\omega_{p}(\pi\bot)=7.48\,10^{15}$ , which yields the line in Fig. 5. The accuracy of the measurement does not warrant a better fit with a model including only 2 pairs of oscillators. This one already demonstrates the respective roles of the model parameters. Now, using the parameters of Tab. 2, we can also deduce the corresponding band intensity, $Q_{abs}/a$. This is displayed in Fig. 6, together with the accompanying Drude fit: $\lambda_{0}=2210$ \AA{\ };  
$\gamma=1.3\,\mu$m$^{-1}$. Thus, this laboratory material approaches qualification as a qualitative model, although comparison with fig. 4 shows that its feature is much weaker than that of the hypothetical model of Sec. 3.

A better laboratory analog might be obtained by trying to mimic more closely the evolutionary trend of carbon grains in the sky. The starting material could be kerogen, whose disordered structure certainly makes it a non-graphitizing carbon. A subsequent slow and protracted high temperature treatment at low ambient pressure would expel all volatile atoms, leaving a nearly pure carbon substance approaching the ideal turbostratic graphite.

Still another analogy can be found in the insoluble organic matter (IOM) found in carbonaceous chondrites extracted from meteorites like Orgueil and Murchison. A large fraction of the  carbon in this material is found in weakly organized aromatic moieties containing about 30 to 150 C atoms, whose average structural parameters have recently been measured by Transmission Electron Microscopy (see Derenne et al. \cite{der}). The average ``diameter" of these basic structural units corresponds to 2 to 3 aromatic rings per layer, with 2 to 3  stacked layers per unit. This is reminiscent of the graphitic chunks constituting our model, Sec. 3. Interestingly, the average interlayer spacing in measured meteoritic stacks is $\sim$0.48 nm, much larger than in terrestrial graphite, suggesting differences in dielectric properties.

\section{Discussion}

The properties of the Fr\"ohlich resonance (Sec. 2) and the sample calculations of Sec. 3 show that the proposed model is able to mimic the UV extinction features, over the observed range of position and width, using dielectric properties that differ only slightly from those of graphite. The differences occur in a limited spectral range, in the blue wing of the $\pi$ resonance, where the measurements of $\epsilon$ are themselves affected by considerable experimental uncertainty. Provided the particle size is no larger than $\sim$100 \AA{\ } for the Rayleigh approximation to be valid, the central wavelength remains within the observational limits, as the observed range of width of the astronomical feature is covered by adjusting only the plasma frequency and the damping constant of only the $\pi\bot$ resonance (in fact, only the wings of this resonance).  We thus concur with the early conclusion of Draine and Malhotra \cite{dm93} that the causes of variation of the IS feature must lie in the dielectric properties of the carrier. These are known to be intimately linked to the details of the solid structure: plasmon-phonon interaction, irregularity of the microscopic ``graphite bricks" (diameter, interlayer distance, etc; see end of Sec. 4), presence of impurities (such as hydrogen), varying strain at interface between bricks, size of the latter, etc (see, for instance, Hecht \cite{hec}, Sorrell \cite{sor}). All of these depend heavily on the previous thermal history of the material (Franklin \cite{fra}, Mrozowski \cite{mro}, Robertson \cite{rob}).² The detailed investigation of the effects of these factors is outside the scope of this paper. However, the particular oscillator parameters listed above, together with the laboratory availability of a close analog, PG, should help defining further the nature of the proposed band carrier. 
In the following, we discuss the adequacy of the model in more detail.

\subsection{Carbon availability}

Let $N(C)$ and $N(H)$ be the column densities of C and H, respectively, towards a sample star, and $\rho$, the grain specific gravity. Take

$\frac{A_{V}}{E_{B-V}}=3.1$ and $N_{H}=5\,10^{21}\,E_{B-V}$ atoms.cm$^{-2}$.mag$^{-1}$,

and assume that, on average, the optical thickness through the line of sight is $\tau(UV)=A_{V}$, then

$\frac{N(C)}{N(H)}=27\frac{\rho}{Q/a}$.

For graphite, $\rho\sim2$ g.cm$^{-3}$, and from Fig. 2, $Q/a$ is of order $10^{6}$, assuming small grains (less than about 100 \AA{\ } in size), the fraction of cosmic carbon required to be in the model grains turns out to be about 54 ppm, or nearly 1/4-1/6 of the available carbon (e.g. Snow and Witt \cite{sno}). Of course the largest fraction must be reserved to less evolved grains.

\subsection{Spurious features}

In agreement with astronomical observations (see Fitzpatrick \cite{fit}) and due to the absence of any alien atom in the ideal model, and to its disordered structure, there should be no features in its spectrum other than the $\pi$ and $\sigma$ resonances overlying a continuum; that is in fact the case for the measured spectrum of polycrystalline graphite. It is apparent, from the figures above, that the continuum is negligible in intensity as compared with the feature itself. It should therefore make a negligible contribution to that which is observed in astronomical spectra.

On the other hand, the $\sigma$ resonance is much stronger than the $\pi$ resonance, and should, therefore, be an important component of the far UV rise beyond 6 $\mu$m$^{-1}$. Indeed, even when the general ascending slope of celestial UV spectra is weak, the local upward bend around 8 $\mu$m$^{-1}$ remains strong (see the large number of diverse spectra collected in Fitzpatrick \cite{fit}). Although available astronomical data do not extend far enough in the UV, they carry no sign of incompatibility with the FUV rise in the measured PG spectrum.

The compact structure of our model drastically reduces the number of surface dangling carbon bonds, and therefore, makes it very difficult for adatoms to stick to the grain. This, however, remains  a subject of further investigation into the possible effects of such adatoms on the profile of the 
$\pi$ and $\sigma$ resonances.

Finally, the perfect isotropy of our model ensures that it causes no polarization of the scattered light, provided the grains are spherical or, in any case, poorly aligned.

\subsection{Relation to fullerene models} 

In the light of our work, the good fit provided by fullerenes and BOs to the (widest) IS features (see Introduction) stems from the fact that, in these particles, the electric dipoles associated with the carbon rings are uniformly distributed over the spherical surface, and their directions are thus uniformly distributed in angle, exactly as assumed in an ideal PG, but on a larger scale. The main problems of the BO model are the electronic spectral structure superposed upon the feature, and the large feature width (1.2-1.6 $\mu$m$^{-1}$; Chhowalla et al. \cite{chh}, Ruiz et al. \cite{rui}).

\subsection{Why Bruggeman ?}
The proposed model being a mixture of components, the choice of a mixing formula is inescapable. The number of available mixing formulae (see Bohren and Huffman \cite{boh}, Stroud \cite{str}) is witness to the theoretical difficulty of the choice. Of course, the arithmetic  weighted mean is simplest. However, in matters dielectric, it is not clear what is the electric property to be averaged: e.g. resistivities cannot be used in a weighted sum. Moreover, common knowledge seems to restrain the use of this ansatz to the inclusion of an impurity in an otherwise homogeneous medium. For comparison, we have also used this averaging: essentially, it shifts the feature of interest to the red by about 100 \AA{\ }. We have also used the Maxwell Garnett mixing formula, which stems from the same basic treatment as Bruggeman's but results from a different approximation (Stroud \cite{str}). The band obtained in this way is shifted, widened and distorted. This does not condemn the mixing formula, but calls attention to possible microresonances already observed experimentally (see Barker \cite{bar}). The Bruggeman formula gives a more acceptable feature shape, at the price of a heavier computation code. Our conclusions regarding the factors governing the feature profile and location are not heavily dependent on this choice. It should be interesting to compare the results with those of the Discrete Dipole Approximation (DDA) applied to a random distribution of dipoles (Draine \cite{dra88}).

Another inconvenience with non-linear formulae like Bruggeman's, is the uncertainty as to whether they satisfy the Kramers-Kronig relation. We have checked this is the case for a mixture of the parallel and perpendicular dielectric functions of graphite (Fig. 1), and, more generally, for a sum of two Lorentzian resonators (Sec. 3 and 4). 

\subsection{Effects of grain shape}

The Fr\"ohlich formula for $Q_{abs}$ and the Bruggeman formula for the average dielectric function  (see Zeng et al. \cite {zen}) both include the ellipsoidal characteristic parameter, $L$, which was set above at 1/3, assuming a spherical shape for both the grain and its component subgrains. As noted by Bohren and Huffman \cite{boh}, shape effects are no weaker in the Rayleigh approximation than for large particles, and the band shape of discs or needles are very different  from the ``Drude" profile. Fortunately, the physics of carbon grain formation and evolution apparently does not allow the evolution towards such extreme shapes. Besides, the grain surface is certainly very irregular and may not be subjected to the same theoretical treatment as perfectly smooth ellipsoids. Finally, even though the term ``surface mode" is usually applied to the Fr\"ohlich resonance, it is clear that the whole grain volume is involved, so the surface shape may not be so important, after all. In order to substantiate these conjectures, we also computed $Q_{abs}$ for values of $L$ symmetrically bracketing 1/3 , from 0.26 to 0.4. It was found that, although the peak wavelength of the feature decreases from 2315 to 2140 \AA{\ }, the average feature still very nearly coincides with the curve of Fig. 4. This is due to the symmetry of the variations, a consequence of the implicit assumption that the fluctuations of L around 1/3 are limited and symmetric.

\section{Conclusion}

Small spheroidal particles made up of a homogeneous mixture of randomly oriented sub-micron chunks of pure graphite have been shown, above, to display a strong Fr\"olich resonance peaking near 2175 \AA{\ }, as does the UV IS extinction feature, the only difference being its slightly larger width. A mild, coherent, tailoring of the dielectric function of graphite around 2000 \AA{\ } suffices to reduce this width to the astronomically observed average. The physical justification of such tailoring lies in   
subtle changes in the damping constant of the $\pi\bot$ resonance, that are likely to result from small alterations of the corresponding electronic band structure of graphite in space. These may be due to different grain formation histories and/or different IS irradiations and other heating effects. Such variations allow the model feature width to cover the range observed in the sky without notably impacting the peak wavelength. The Discussion shows that none of the usual observational constraints disagrees with this model.

\section{Acknowledgments}
We thank the reviewer for useful suggestions, and Dr J.-M. Perrin (Observatoire de Haute-Provence) for useful documents.

\vfill\eject

Note added in proof:

The property of separate dependencies of the central wavelength and the FWHM of the Fr\"ohlich resonance (Sec. 2, eq. 4 and 6) was deduced from the crude approximation in eq. 3. If, instead, one goes through the tedious mathematics of differentiating $Q(\omega)$ to determine the frequencies at the peak and at half-maximum, one obtains the following, more accurate relations, to first order in $\gamma/\omega$, in the relevant range of the parameters:
\begin{eqnarray}
\omega_{m}=\omega_{0}+\frac{\omega_{p}^{2}}{6\,\omega_{0}}\\
FWHM=\gamma\\
\frac{\epsilon_{1m}+2}{\epsilon_{2m}}=\frac{\epsilon_{2m}}{3}=\frac{3\omega_{0}\gamma}{\omega_{p}^{2}}
\end{eqnarray}

\end{document}